\documentclass[aps,prl,reprint,amsmath,amssymb,superscriptaddress]{revtex4-1}

\usepackage{graphicx}
\usepackage{dcolumn}
\usepackage{bm}
\usepackage{xcolor}
\usepackage[english]{babel}
\usepackage[hidelinks]{hyperref}

\newcommand{\rev}[1]{\textcolor{black}{#1}}

\begin{document}

\title{Spatiotemporal shaping of attosecond X-rays with time-dependent orbital angular momentum}

\author{Chenzhi Xu}
\affiliation{Shanghai Institute of Applied Physics, Chinese Academy of Sciences, Shanghai 201800, China}
\affiliation{University of Chinese Academy of Sciences, Beijing 100049, China}
\affiliation{European XFEL, 22869 Schenefeld, Germany}

\author{Jiawei Yan}
\email{jiawei.yan@desy.de}
\affiliation{Deutsches Elektronen-Synchrotron DESY, 22603 Hamburg, Germany}

\author{Gianluca Geloni}
\email{gianluca.geloni@xfel.eu}
\affiliation{European XFEL, 22869 Schenefeld, Germany}

\author{Christoph Lechner}
\affiliation{European XFEL, 22869 Schenefeld, Germany}

\author{Haixiao Deng}
\email{denghx@sari.ac.cn}
\affiliation{Shanghai Advanced Research Institute, Chinese Academy of Sciences, Shanghai 201210, China}

\begin{abstract}

\rev{Attosecond X-ray pulses are indispensable for exploring ultrafast phenomena in matter on \AA{}ngstrom and attosecond scales. Here we propose a new method to realize spatiotemporal shaping of attosecond X-rays through temporal control of the orbital angular momentum mode content using an X-ray free-electron laser. The method exploits transverse-mode-dependent frequency pulling in a deliberately detuned second stage, together with slippage between the seed and the amplified radiation. Three-dimensional simulations show a double-spike waveform in which the two spikes carry different dominant topological charges. The spike separation is tunable and can reach several hundred attoseconds. This provides a source-level route to spatiotemporally structured attosecond X-rays with controllable temporal structure and topological mode content.}

\end{abstract}

\maketitle

Since the first experimental generation and measurement of attosecond light pulses~\cite{paul2001observation,hentschel2001attosecond}, attosecond science has undergone rapid development. The ability to observe phenomena on attosecond timescales has revolutionized our understanding of ultrafast electron dynamics~\cite{krausz2009attosecond}. Over the past two decades, the generation of attosecond light pulses has relied mainly on a process known as high-harmonic generation (HHG)~\cite{paul2001observation,mairesse2003attosecond,li2020attosecond}. However, a great challenge arises from the wavelength scaling inherent in the HHG process, which leads to a dramatic decrease in conversion efficiency with increasing X-ray photon energy. This limitation severely prevents our ability to investigate matter at \AA{}ngstrom--attosecond resolutions, which are typical spatiotemporal scales for electron dynamics in various materials and ultrafast phenomena.

X-ray free-electron lasers (XFELs), the new generation of particle accelerator--based light sources, generate extremely bright X-ray pulses at \AA{}ngstrom wavelengths using relativistic electron beams in a long undulator~\cite{emma2010first,fel3,RevModPhys.88.015006, fel4,fel5,fel6,huang2021features}. Recent advancements in XFELs have made it possible to generate high-power attosecond pulses at both soft and hard X-ray wavelengths~\cite{huang2017generating,duris2020tunable,franz2024terawatt,yan2024terawatt}. Moreover, attosecond XFELs could generate attosecond pairs~\cite{duris2020tunable,guo2024experimental} with a separation of only a few tens to a few hundred attoseconds. Such attosecond pulse pairs pave the way for time-resolved studies of electron dynamics with unprecedented structural resolution. The first such experiment in the soft X-ray regime has been reported, focusing on the electronic response following valence ionization in liquid water~\cite{li2024attosecond}. 

In parallel, the generation of attosecond X-ray pulses carrying orbital angular momentum (OAM) has garnered significant attention. OAM light~\cite{allen1992orbital} is characterized by a unique phase structure, which can be mathematically expressed as $\exp(im\phi)$, where $\phi$ represents the azimuthal angle, and $m$ denotes the topological charge~\rev{(topological quantum number)}. X-ray pulses carrying OAM are promising means to trigger new phenomena through light--matter interaction, such as the exploration of chirality~\cite{fanciulli2021electromagnetic,fanciulli2022observation,mccarter2023antiferromagnetic, fanciulli2025Magnetic}, photoionization experiments~\cite{picon2010photoionization,DeNinno2020PhotoelectricTwist}, resonant X-ray inelastic scattering~\cite{rury2013examining}, and proposals for probing ultrafast electronic coherences in molecular systems~\cite{yong2022direct,tang2024pushing}.~\rev{Beyond isolated attosecond OAM pulses, spatiotemporally structured light featuring time-varying OAM has attracted increasing attention in recent years~\cite{doi:10.1126/science.aaw9486,Sun:23,Dong:24,Liu2025UltrafastBurstsSTVP}. The central idea is the joint control of the temporal structure and the transverse topological degree of freedom.}

While experimental demonstration remains forthcoming, generating attosecond XFEL pulses carrying OAM is envisioned through direct shaping techniques using optical components~\cite{peele2002observation,sakdinawat2007soft,vila2014characterization,seiboth2019refractive,lee2019laguerre}, such as spiral phase plates (SPP)~\cite{oemrawsingh2004production} or spiral Fresnel zone plates~\cite{vila2014characterization,ribivc2017extreme}, the utilization of helical undulators~\cite{hemsing2009helical,hemsing2011generating,bahrdt2013first,hemsing2013coherent,hemsing2014first,hemsing2020coherent,morgan2022x,Morgan2025Poincare}, externally seeded FEL schemes~\cite{hemsing2012echo,PhysRevLett.112.203602}, XFEL oscillator~\cite{huang2021generating, huang2024rapidly}, and self-seeded FEL setups~\cite{yan2023self,PhysRevLett.134.205001}.~\rev{However, despite these approaches toward OAM-carrying X-ray pulses, extending temporal--topological control to the attosecond XFEL regime remains largely unexplored and nontrivial. In this context, our work introduces an XFEL-based route to generate attosecond X-ray pulse pairs whose two temporal spikes carry distinct topological charges, enabling concurrent control of the temporal structure and the OAM mode content at X-ray wavelengths.}

\begin{figure*}
    \centering
    \includegraphics[width=18cm]{scheme_v8_1.jpg}
    \caption{Schematic of the generation of attosecond X-ray pulse pairs with different dominant topological charges. Temporal power of the FEL pulse at (a) the end of the first stage, (b) the entrance of the second stage and (c) the end of the second stage. The electron-beam head is on the left-hand side of each panel. }
    \label{fig:1scheme}
\end{figure*}

The proposed scheme is based on a self-seeded FEL setup, as illustrated in Figure~\ref{fig:1scheme}. The first stage undulator is utilized to generate an attosecond XFEL seed. The attosecond seed can be obtained by modulating the electron beam with a few-cycle laser pulse, such as the enhanced self-amplified spontaneous emission (ESASE) scheme~\cite{zholents2005method}, or via collective effects induced by the electron beam itself~\cite{yan2024terawatt,yan2025attoshine}. An SPP, or other optical components, then imprints a helical phase onto the seed, creating an attosecond pulse carrying OAM. A small chicane between the stages removes microbunching and delays the beam, ensuring overlap of the OAM seed with a fresh electron slice. This requires a long electron beam with modulation confined to the tail. The modified seed enters the second undulator to generate attosecond pulse pairs with distinct topological charges. 

During a typical FEL process, the spontaneous emission comprises multiple transverse modes, while the fundamental mode with the highest gain due to better beam overlap, gradually dominates~\cite{kim2017synchrotron,moore1985high,scharlemann1985optical}. However, higher-order transverse modes can remain present, especially far from saturation. Thus, the seed field at the exit of the first stage can be regarded as a superposition of azimuthal modes proportional to $\exp(-i n \phi)$, where $n$ is the azimuthal index. When this field passes through a helical optical element such as an SPP, a phase factor $\exp\!\left(i l \phi\right)$ is applied, shifting the topological charge of each mode by $l$, so that the output topological charge becomes $m = l - n$. At the entrance of the second-stage undulator, the slowly varying field envelope in the frequency domain can then be written as
\begin{equation}
\widetilde{E}(\hat{r},\phi,\hat{\omega})
= \sum_{m=-\infty}^{\infty} A^{(m)}(\hat{\omega})\, \Phi_{m}(\hat{r},\hat{\omega}) \exp\!\left(i m \phi\right),
\label{eqafterspp}
\end{equation}
where $\hat{r}$ is the normalized transverse coordinate, $\hat{\omega}$ is the normalized angular frequency, $A^{(m)}(\hat{\omega})$ are complex spectral amplitudes, and $\Phi_{m}(\hat{r},\hat{\omega})$ denote the corresponding transverse mode profiles~\cite{yan2023self,saldin2000diffraction}. In our configuration with $l = 1$, the dominant $n = 0$ component before the SPP is mapped to the $m = 1$ mode, while a small residual $m = 0$ component is also present.

In the second stage, the growth of different modes of the field is governed by the spectral overlap between the attosecond seed spectrum and the FEL gain spectrum. When the seed spectrum overlaps with the FEL gain spectrum, the attosecond OAM pulse is significantly amplified, as discussed in~\cite{yan2023self,xu2025generation}. However, when the undulator is strongly detuned and the gain spectrum only partially overlaps with the spectrum of the seed pulse, a second radiation spike emerges following the seed pulse. This temporal separation arises because, in the large-detuning regime, the weak interaction between the seed and the electron beam causes the seed pulse to slip ahead at the speed of light~\cite{robles2025spectrotemporal}, while the radiation generated through FEL gain propagates with a slower group velocity, given by $v_g = c \!\left( 1 - \frac{2\lambda_r}{3\lambda_u} \right)$. In addition to its temporal delay, the second spike exhibits distinct spectral features. The central wavelength of this second spike is determined by the interplay between the seed spectrum and the gain in the undulator. This shift in the emission wavelength, caused by their spectral interaction, is a phenomenon known as frequency pulling~\cite{haixiao2008short,allaria2010tunability,allaria2011experimental,robles2023attosecond}. Owing to the mode-dependent gain response to frequency pulling, tuning the detuning of the second-stage undulator provides a handle to shift the pulled central wavelength and thereby tailor the second spike. For convenience, we define the normalized detuning as $\hat{C} = \frac{k_u \delta \omega}{\Gamma \omega_1}$, as in Ref.~\cite{saldin1999physics}, where $\omega_1$ is the FEL resonant frequency of the first-stage undulator, $\Gamma$ is the peak gain wavenumber, $k_u$ is the undulator wavenumber, and $\delta \omega$ denotes the frequency difference between the resonances of the two stages.

To investigate how second-stage detuning influences the evolution of FEL pulses with different transverse modes, we perform numerical simulations using Genesis 1.3 version 4~\cite{reiche1999genesis,reiche2022genesis,reiche2024status}. The initial electron shot noise is disabled in the simulation. For further study, a photon energy of $6\ \mathrm{keV}$, an electron beam energy of $14\ \mathrm{GeV}$, a peak current of $5\ \mathrm{kA}$, a normalized emittance of $0.5 \ \mathrm{mm}$ mrad, an average $\beta$ function of $32\ \mathrm{m}$, a slice energy spread of $3\ \mathrm{MeV}$, an undulator period of $40\ \mathrm{mm}$ and an undulator segment length of $ 5\ \mathrm{m}$ are considered, consistent with the hard X-ray undulators at the European XFEL~\cite{fel5}.  

\begin{figure}[!htb]
	\centering	
	\includegraphics[width=1\linewidth]{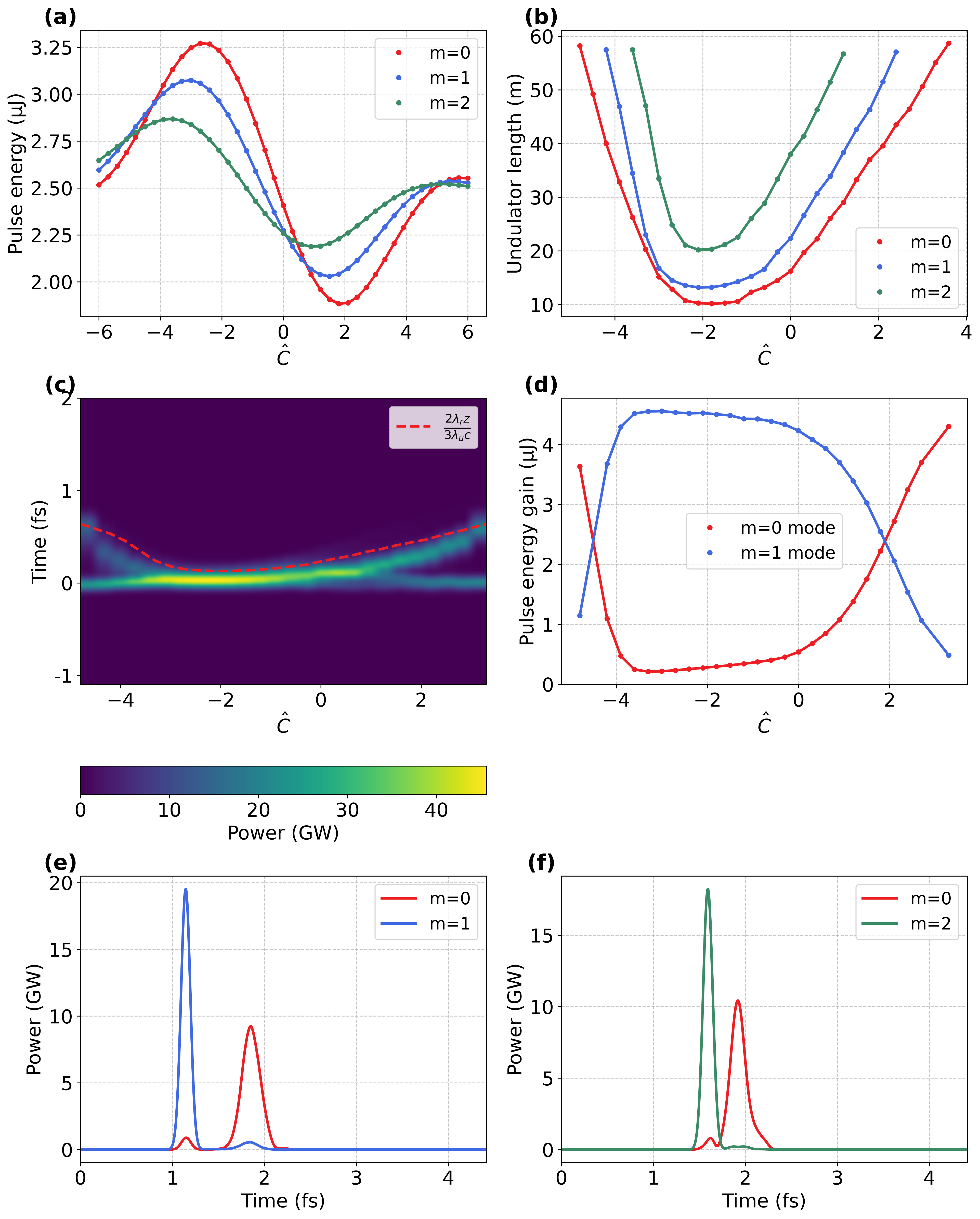}
	\caption{Simulation results under various detuning conditions.  
(a) and (b) show results for pure mode seed pulses.  
(a) Pulse energy at the exit of the first undulator segment for different transverse modes. (b) Undulator length required for the pulse energy to increase by a factor of $e$.  
(c)–(f) show results for mixed mode seed pulses.  
(c) Power profile at the position where the pulse energy has increased by a factor of $e$.   
(d) Energy gain of different modes as a function of reduced detuning.  
(e) Temporal power profile at the exit of the twelfth undulator segment for $\hat{C}=4.2$.  
(f) Temporal power profile for $m=0$ and $m=2$ mode at the exit of the seventh undulator segment for $\hat{C}=2.0$.}
	\label{fig:2}
\end{figure}

At the entrance of the undulator, three types of seed pulses with different transverse modes (topological charges $0$, $1$, and $2$) are considered, each having a peak power of $20\ \mathrm{GW}$ and a pulse duration of $118\ \mathrm{as}$.~\rev{They have identical radial dependence and temporal envelope. In the simulations, these idealized seed pulses are first propagated through a $2\ \mathrm{m}$ free-space drift before reaching the undulator entrance, consistent with the experimental setup.} Figure~\ref{fig:2}(a) shows the pulse energy at the exit of the first undulator segment as a function of the detuning parameter for different transverse modes. This transverse mode frequency pulling effect in the low-gain regime has been studied in~\cite{benson1985transverse, huang2021generating,huang2024rapidly}. Furthermore, Figure~\ref{fig:2}(b) shows the gain length for each mode, revealing that the fundamental mode grows faster near resonance, while all modes require much longer gain lengths under strong detuning. This temporal stretching enables the formation of well-separated attosecond pulse pairs~\cite{yang2020postsaturation}. 

To illustrate the generation of twin attosecond pulses, we further consider an attosecond seed pulse comprising a mixture of transverse modes, while keeping its peak power and pulse duration unchanged. Specifically, we assume a superposition of the $m = 0$ (4\%) and $m = 1$ (96\%) modes as an example. Figure~\ref{fig:2}(c) presents the power profile as a function of the reduced detuning parameter $\hat C$ at the point where the pulse energy has increased by a factor of $e$. Time $t=0$ corresponds to the position of the pulse propagating at the speed of light. We can see that when the detuning is sufficiently large, there are two spikes that coexist due to the difference in group velocities. The temporal separation between them at this position is $\Delta t = \frac{2 \lambda_r z}{3 \lambda_u c}$, as indicated by the lines in Figure~\ref{fig:2}(c). The dependence of gain on topological charge under different detuning values is illustrated in Figure~\ref{fig:2}(d). Notably, the gain of the $m = 0$ mode surpasses that of the $m = 1$ mode when $\hat{C} > 1.9$ or $\hat{C} < -4.5$, due to the shorter gain length of the fundamental mode under strong detuning. Consequently, even a weak $m=0$ component in the seed can dominate the second spike. For example, at $\hat{C} = 4.2$, twin attosecond pulses appear at the end of the 12th undulator segment, with the leading spike carrying $m = 1$ and the trailing spike $m = 0$ (Figure~\ref{fig:2}(e)). Similarly, a seed composed of 96\% $m=2$ and 4\% $m=0$ yields pulse pairs at $\hat{C} = 2.0$ (Figure~\ref{fig:2}(f)). Other combinations, such as $m = 1$ and $m = 2$, can also be generated by appropriately tailoring the seed composition and detuning.

Building on the previous analysis, we further examine attosecond pulse pair generation under more realistic conditions, where the initial transverse mode composition varies due to electron shot noise. In this case, we employ the ESASE scheme to generate a $6\ \mathrm{keV}$ attosecond seed in the first stage, using the same beam and undulator parameters described earlier.~\rev{A short wiggler is used to impose an energy modulation on the electron beam, applied only to the trailing part of the bunch. The tail is modulated by a 1030~nm laser pulse with 4~mJ energy and a 4~fs (FWHM) duration in a two-period wiggler with a 0.7~m period. A downstream dispersive section with $R_{56}=41~\mu$m converts the energy modulation into a density modulation, yielding a strong current spike with a peak current of about 12~kA~\cite{supp}\nocite{yan2022simulation}.} The first stage uses nine undulator segments with a taper of $\Delta K/K = 0.11\%$ starting at the third segment, yielding a FEL pulse with an energy of $3.73\ \mu\mathrm{J}$ and a duration of $132\ \mathrm{as}$, dominated by the fundamental mode ($92.8\%$ of the total energy), as shown in Figure~\ref{fig:1scheme}(a). An SPP is then introduced to impose a helical phase factor $\exp(i l \phi)$ with $l=1$, assuming no change in pulse intensity after transmission. Upon entering the second-stage amplifier, the pulse is primarily composed of the $m=1$ mode, with a fraction of $92.8\%$, while the $m=0$ mode constitutes $4.11\%$, as shown in Figure~\ref{fig:1scheme}(b). 

\begin{figure}[!htb]
	\centering	
	\includegraphics[width=1\linewidth]{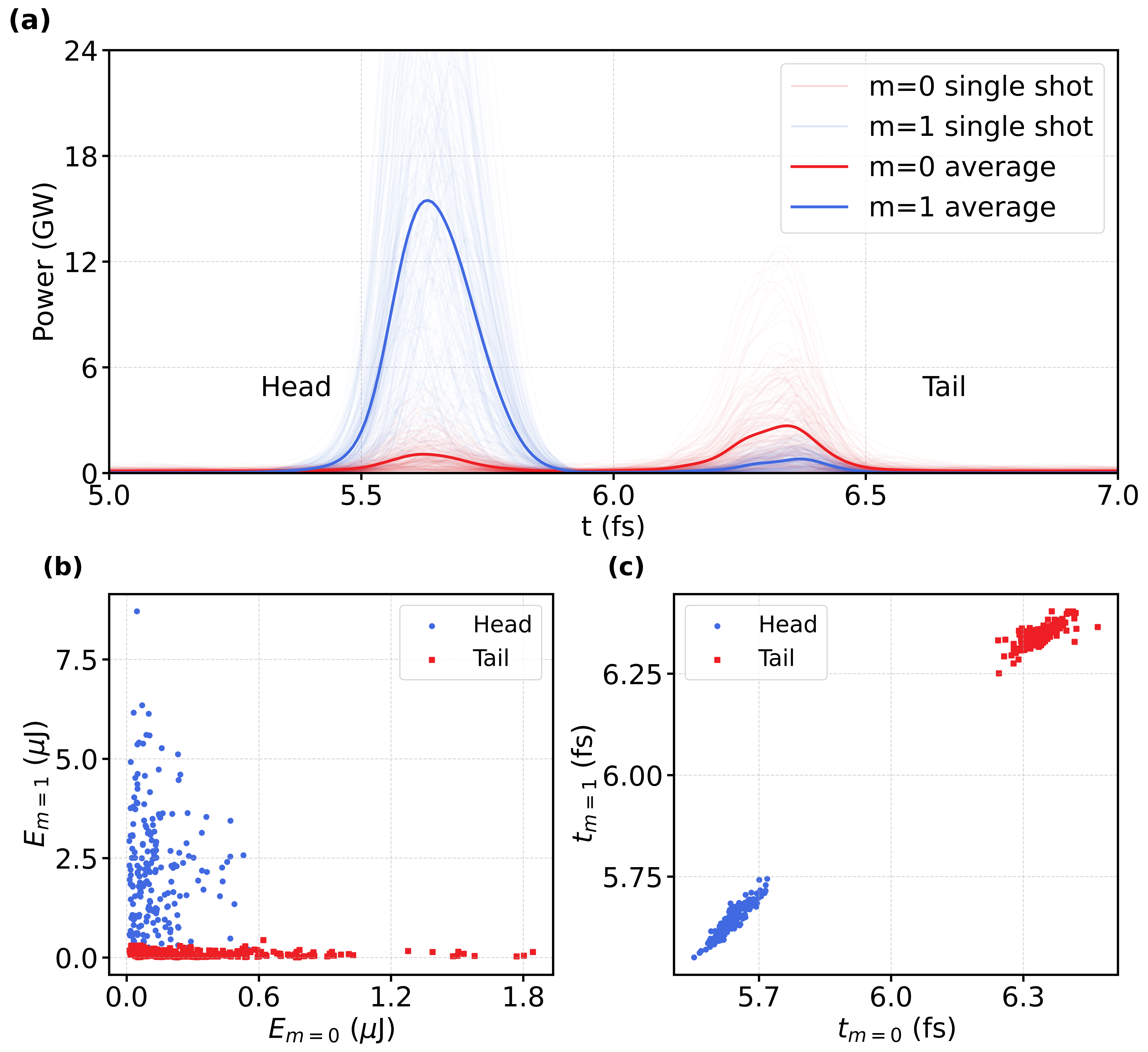}
	\caption{(a) Temporal power at the end of the second stage for 200 simulation runs. Thin lines show individual single-shot realizations, and thick lines show the ensemble-averaged power for the $m=0$ (red) and $m=1$ (blue) components. (b) Scatterplot of $E_{m=1}$ versus $E_{m=0}$ for the head (blue) and tail (red) spikes, where the pulse energies are evaluated within the FWHM window of each spike. (c) Scatterplot of $t_{m=1}$ versus $t_{m=0}$ for the head (blue) and tail (red) spikes, where the arrival time is defined as the intensity-weighted temporal centroid within the same window.}
	\label{fig:5}
\end{figure}

\rev{To generate double attosecond X-ray pulses with distinct topological charges, we use a chicane to delay the electron beam by $4~\mu\mathrm{m}$, thereby arranging temporal overlap between the attosecond OAM seed and a fresh slice of the bunch.} The second-stage undulator consists of eight segments and is operated at an optimized detuning of  $\hat{C}=7.1$, with a taper of  $\Delta K/K = -0.04\%$. As a result, double attosecond X-ray pulses are produced at the exit of the second-stage undulator, as shown in Figure~\ref{fig:1scheme}(c). The radiation exhibits two clearly separated peaks. The trailing peak is predominantly composed of the  $m=0$ mode, accounting for 94.2\% of the energy, with a peak power of $6.7\ \mathrm{GW}$ and a FWHM pulse duration of $158\ \mathrm{as}$. The leading peak is dominated by the  $m=1$ mode with a 91.3\% contribution, a peak power of $22.1\ \mathrm{GW}$, and a FWHM pulse duration of $142\ \mathrm{as}$. The temporal separation between the two peaks is $736\ \mathrm{as}$. To assess the stability of this scheme, we perform 200-shot statistical simulations including random shot noise. The ensemble results are shown in Figure~\ref{fig:5}. On average, the trailing spike has a peak power of $3.5\ \mathrm{GW}$, with the $m=0$ mode contributing 77\% of the energy in that window. The leading spike has an average peak power of $17.7\ \mathrm{GW}$, with the $m=1$ mode accounting for 87\%. The average temporal separation between the spikes is $719\ \mathrm{as}$. These results confirm the robustness of the scheme, even in the presence of shot-to-shot fluctuations due to initial electron noise.~\rev{In addition, using the same scheme with a different parameter set, attosecond OAM pulses with peak power exceeding 300~GW can be obtained~\cite{supp}. }

\begin{figure}[!htb]
	\centering	
	\includegraphics[width=1\linewidth]{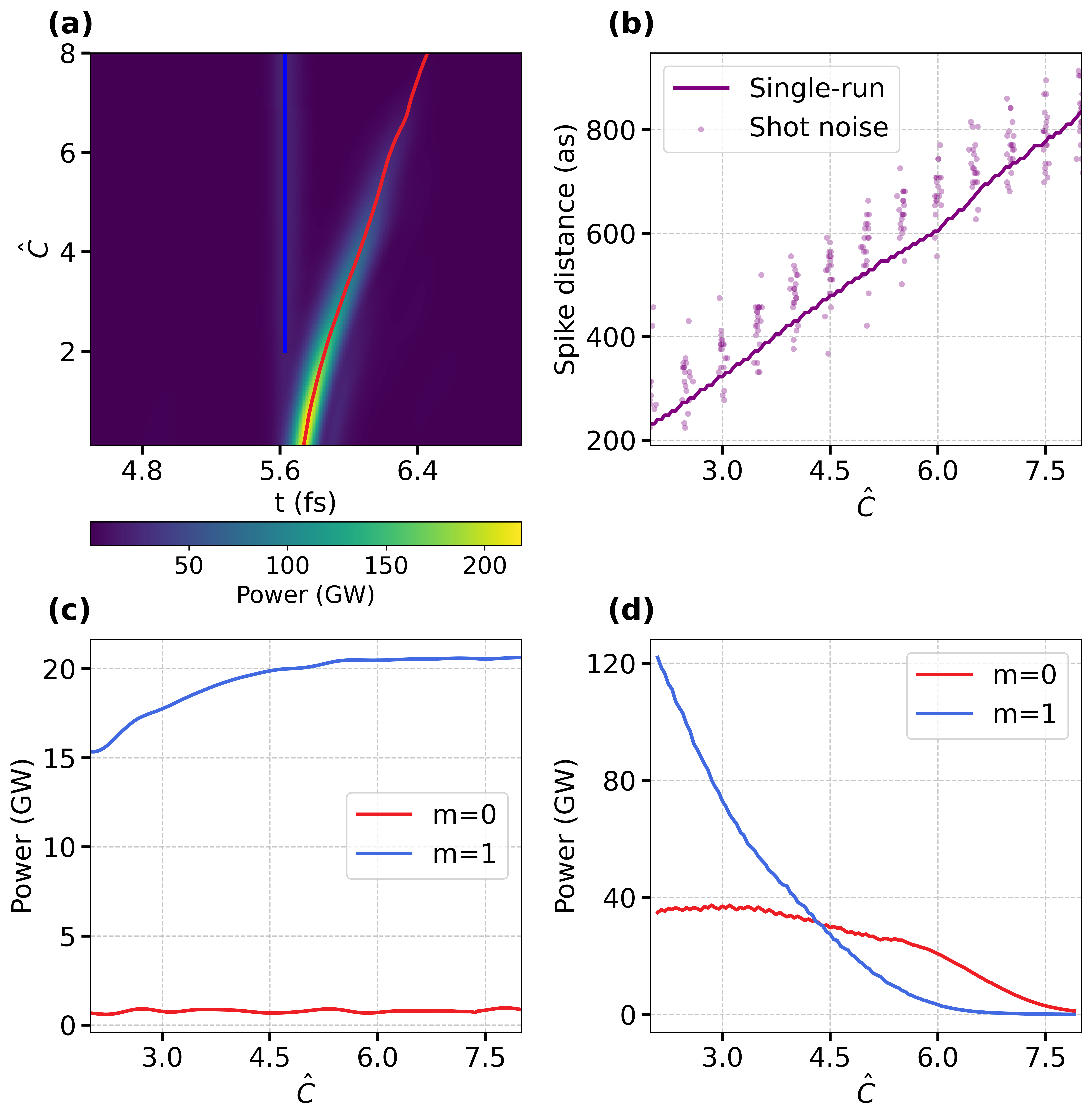}
	\caption{(a) Temporal power of the second-stage radiation as a function of the reduced detuning parameter $\hat{C}$ for a representative single-run simulation. The blue and red lines trace the tail and head spikes, respectively. (b) Temporal separation between the two spikes as a function of $\hat{C}$. The solid line shows a representative single-run result, while the points indicate the statistical spread obtained from 20 shot-noise simulation runs at each value of $\hat{C}$. (c),(d) Single-run power of the $m=0$ (red) and $m=1$ (blue) components sampled at the positions of the head spike (c) and the tail spike (d), respectively. In panels (b)–(d), only values of $\hat{C}$ for which two distinct spikes can be clearly resolved are shown.}
	\label{fig:6}
\end{figure}

To further investigate the influence of detuning, we maintain the tapering and other second-stage parameters while varying the reduced detuning $\hat{C}$ from 0.0 to 8.0. As shown in Figure~\ref{fig:6}(b), the temporal separation between the two peaks decreases as the detuning decreases. The separation reaches $827\ \mathrm{as}$ at $\hat{C}=8.0$, drops to $231\ \mathrm{as}$ at $\hat{C}=2.0$, and vanishes for $\hat{C}<2.0$ as the spikes merge into a single pulse. For $\hat{C}>8.0$, only the seed spike remains. Figure~\ref{fig:6}(d) shows that the $m=1$ component of the trailing spike is negligible at large detuning but grows as $\hat{C}$ decreases, surpassing the $m=0$ component once $\hat{C}$ falls below 4.4. Meanwhile, the relative power of the two spikes can be tuned by varying $\hat{C}$, with an coupling to the temporal delay.~\rev{This detuning-dependent correlation can be viewed as an additional tuning knob when a controlled mixture of transverse mode content and relative spike strength is desired~\cite{Jain2024IntrinsicDichroismHelicalLight}. }

In this Letter, we propose an XFEL-based method for spatiotemporal shaping of attosecond X-rays by generating a double-spike waveform whose two temporal spikes have different dominant topological charges. Three-dimensional simulations show that the separation of leading and trailing attosecond spikes can be tuned in the sub-femtosecond range. Shot-noise simulations indicate that the two-spike structure and the dominant-mode assignment persist under shot-to-shot fluctuations. We also outline an alternative implementation in which the first stage is driven to higher energy while the seed remains nearly free of $m \neq 0$ components, and a controlled $m=0$ fraction is introduced by the SPP while still imprinting the helical phase on the seed \cite{supp}.

More broadly, the proposed approach provides source-level control of both temporal structure and transverse topology, which is challenging to achieve by external optical shaping alone at hard X-ray photon energies. While we present the combination of topological charges $m = 0$ and $m = 1$ as a representative case, other mode combinations such as $m = 0$ and $m = 2$ are also feasible. Furthermore, by introducing an additional SPP after the second stage undulator with higher-order phase profiles, such as $\exp(-i 3\phi)$, it becomes possible to generate pulse pairs carrying even larger topological charges, for instance $m = -3$ and $m = -2$.~\rev{Such structured pulses can further enable ultrafast experiments by providing the topological charge as an additional user-selectable control parameter~\cite{PhysRevLett.98.157401,Rouxel2022HardXrayHelicalDichroism, Fang2022ProbingOAM,Begin2025OAMIonization,Session2025OpticalPumpingVortexLight}.} In the future, this method may be extended to generate pulse pairs with tailored polarization, multi-color spectra, or more exotic spatial mode structures, further expanding its potential in emerging attosecond X-ray science. 

The authors thank Nanshun Huang from SARI, Haiyang Li from SINAP, Giovanni Perosa from European XFEL, Ye Chen, Marc Guetg, Tianyun Long, and Winfried Decking from DESY for helpful discussions. This work was supported by the National Natural Science Foundation of China (12125508, 12541503). Chenzhi Xu thanks the support of CAS-DAAD Joint Scholarship. Jiawei Yan acknowledges support from DESY (Hamburg, Germany), a member of the Helmholtz Association (HGF), and the European XFEL (Schenefeld, Germany). The numerical simulations were supported by the ``MAXWELL'' computational resources operated at Deutsches Elektronen-Synchrotron DESY, Hamburg, Germany.

\bibliography{ssoam}

\appendix

\end{document}


\title{Supplementary Material for ``Spatiotemporal shaping of attosecond X-rays with time-dependent orbital angular momentum''}

\author{Chenzhi Xu}
\affiliation{Shanghai Institute of Applied Physics, Chinese Academy of Sciences, Shanghai 201800, China}
\affiliation{University of Chinese Academy of Sciences, Beijing 100049, China}
\affiliation{European XFEL, 22869 Schenefeld, Germany}

\author{Jiawei Yan}
\email{jiawei.yan@desy.de}
\affiliation{Deutsches Elektronen-Synchrotron DESY, 22603 Hamburg, Germany}

\author{Gianluca Geloni}
\email{gianluca.geloni@xfel.eu}
\affiliation{European XFEL, 22869 Schenefeld, Germany}

\author{Christoph Lechner}
\affiliation{European XFEL, 22869 Schenefeld, Germany}

\author{Haixiao Deng}
\email{denghx@sari.ac.cn}
\affiliation{Shanghai Advanced Research Institute, Chinese Academy of Sciences, Shanghai 201210, China}

\maketitle

\section{Simulation of enhanced self-amplified spontaneous emission} 

\begin{figure}[!htb]
	\centering	
	\includegraphics[width=1\linewidth]{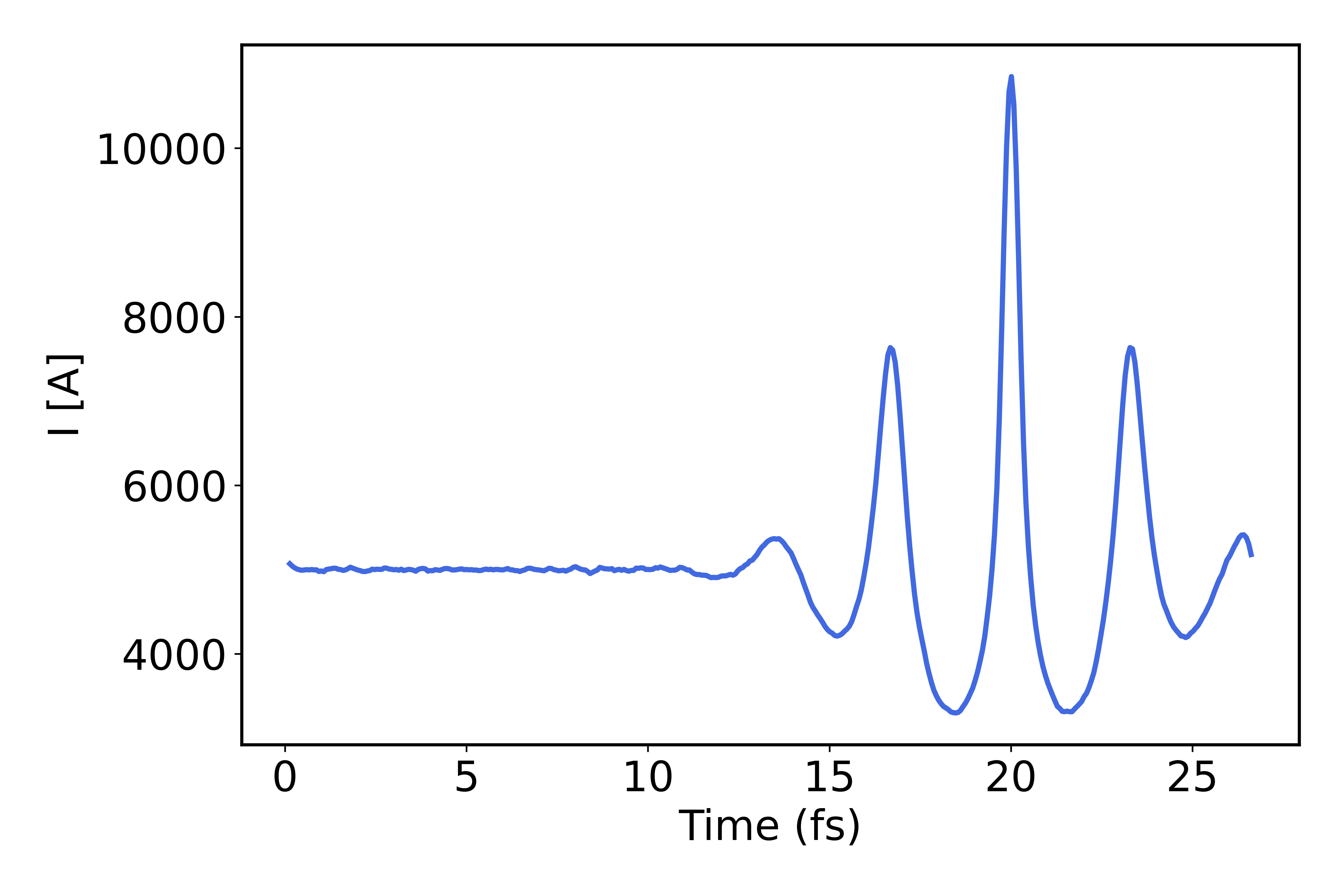}
	\caption{Current profile of the electron bunch after modulation. The left side corresponds to the head of the electron bunch.}
	\label{fig:supply1}
\end{figure}

The simulations in this letter are based on parameters representative of the SASE2 beamline at the European XFEL~\cite{fel5}. The electron beam is energy-modulated via copropagation with a laser pulse in a short wiggler. The laser wavelength is 1030~nm, with a pulse energy of 4~mJ and an FWHM duration of 4~fs. The interaction takes place in a two-period wiggler with a period of 0.7~m. The chicane and arc upstream of the undulator, with $R_{56}=41~\mu\mathrm{m}$, convert the energy modulation into a density modulation~\cite{yan2022simulation}. The electron beam is then well compressed before entering the undulator. The resulting current profile after compression is shown in Fig.~\ref{fig:supply1}. 

\begin{figure}[!htb]
	\centering	
	\includegraphics[width=1\linewidth]{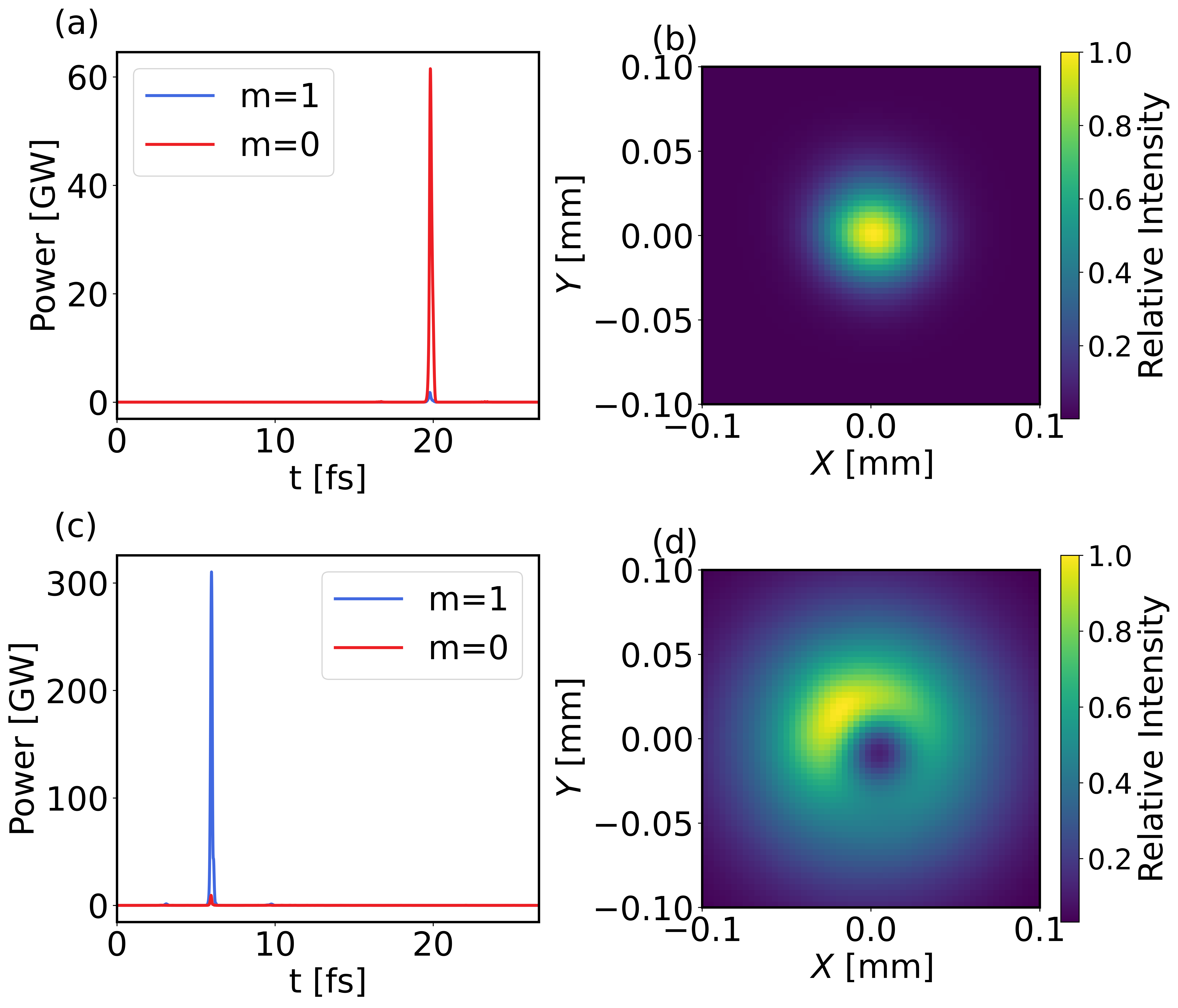}
	\caption{(a) Temporal power profile and (b) transverse profile of the FEL pulse at the end of the first stage. (c) Temporal power profile and (d) transverse profile of the FEL pulse at the end of the second stage.}
	\label{fig:supply2}
\end{figure}

To generate a high-peak-power attosecond pulse with orbital angular momentum (OAM), the energy-modulated beam is sent to the first undulator stage to generate an attosecond seed pulse. Compared with the case of generating attosecond pulse pairs with different topological charges, here ten undulator segments are used in the first stage, while the tapering is kept at $\Delta K/K = 0.11\%$ from the third undulator segment onward. At the end of the first stage, an attosecond seed pulse with a peak power of 65~GW is obtained, as shown in Figs.~\ref{fig:supply2}(a) and \ref{fig:supply2}(b). A chicane is then used to delay the electron beam by $4~\mu\mathrm{m}$ and to wash out the microbunching. The attosecond seed is converted into an attosecond OAM seed by a spiral phase plate (SPP) and is overlapped with a fresh slice of the electron beam. In the second stage, six undulator segments are used without additional detuning to generate attosecond radiation with OAM at 6~keV, and a tapering of $\Delta K/K = -0.04\%$ is applied. At the end of the second stage, the attosecond OAM pulse reaches a peak power of 320~GW with a pulse duration of 115~as, as shown in Figs.~\ref{fig:supply2}(c) and \ref{fig:supply2}(d). 

\section{Time-varying X-ray OAM pulses using a detuned spiral phase plate}

\begin{figure}[t]
	\centering	
	\includegraphics[width=1\linewidth]{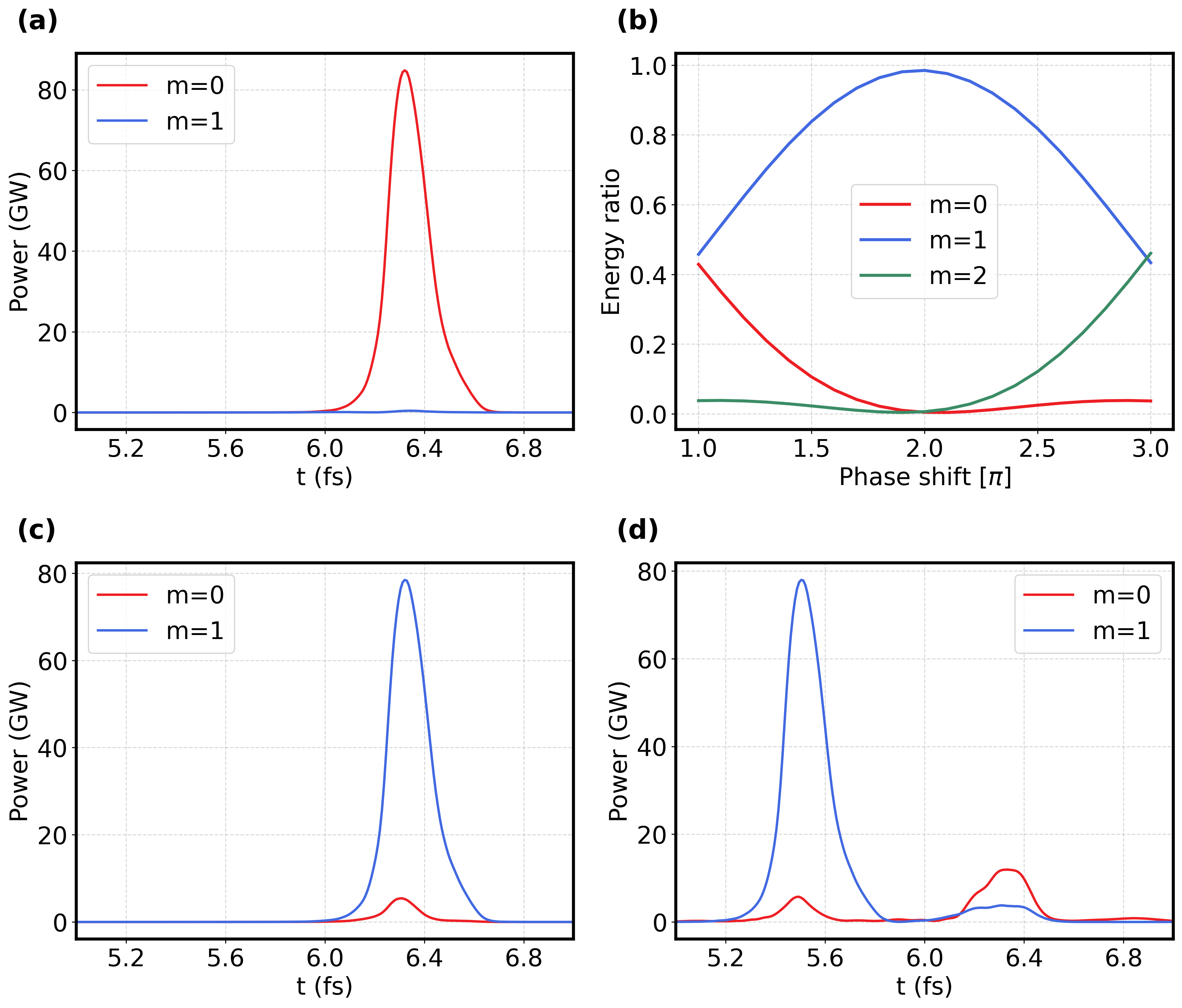}
	\caption{Single-shot example demonstrating the generation of a high-intensity time-varying-OAM X-ray pulse.
(a) Temporal power profiles of the $m=0$ and $m=1$ components at the exit of the first stage.
(b) Energy fraction of different OAM components as a function of the SPP phase shift.
(c) Power profiles at the entrance of the second stage for an SPP phase shift of $1.7\pi$.
(d) Power profiles at the exit of the second stage.}
	\label{fig:supply3}
\end{figure}

Here, we discuss a alternative method for generating high-intensity, time-varying X-ray OAM pulses using a detuned spiral phase plate (SPP). By operating the first-stage amplification close to saturation, the higher-order OAM content of the seed remains relatively weak. In this regime, a detuned SPP can intentionally introduce a controlled $m=0$ component while imprinting the helical phase. 

As an illustrative single-shot example, we consider a first-stage undulator comprising 11 segments and apply a taper of $\Delta K/K = 0.11\%$ starting from the third segment, while keeping all other parameters the same as in the ESASE-based baseline setup of this Letter. At the end of the first stage, we obtain an $m=0$ pulse with a modal purity exceeding 98\% and a peak power above 80~GW, as shown in Fig.~\ref{fig:supply3}(a). By varying the SPP phase shift, as shown in Fig.~\ref{fig:supply3}(b), the $m=0$ energy fraction at the entrance of the second stage can be increased by choosing an appropriate phase. For an SPP phase shift of $1.7\pi$, the power profile at the entrance of the second stage is shown in Fig.~\ref{fig:supply3}(c), where the $m=0$ energy fraction in the main spike is 6.4\%. With the second-stage undulator detuned to $\hat{C}=9.4$ and a taper of $\Delta K/K = -0.04\%$, the power profile at the exit of the second stage is shown in Fig.~\ref{fig:supply3}(d). Under these conditions, a high-intensity time-varying-OAM X-ray pulse can also be generated.

\bibliography{ssoam}